# Tutorial: Opto-mechanical cooling by the back-action of cavity photons

Masud Mansuripur

College of Optical Sciences, The University of Arizona, Tucson, Arizona 85721

**Abstract**. We present a simple classical analysis of light interacting with a Fabry-Perot cavity consisting of a fixed (dielectric) front mirror and a vibrating rear mirror. In the adiabatic approximation, the returning light exhibits sideband symmetry, which will go away once the photon lifetime becomes comparable to or longer than the oscillation period of the rear mirror. When the oscillation period is short compared to the cavity photon lifetime, one must approach the problem differently, treating the vibrating mirror as a scatterer which sends a fraction of the incident light into sideband frequencies. With proper detuning, the cavity's internal radiation pressure could either dampen or amplify the vibrations of the mirror; the former is the physical principle behind opto-mechanical cooling by the back-action of cavity photons.

**1. Introduction**. Many methods of measuring mechanical displacement rely on the coupling between light and mechanical degrees of freedom. The back-action of light in general, and cavity-enhanced radiation pressure, in particular, can dominate the mechanical dynamics of certain opto-mechanical systems which have emerged in recent years as micro- and nano-fabrication techniques have matured. A wealth of information is now available, addressing both theoretical and experimental aspects of cavity opto-mechanics [1-17]. Our modest goal in this tutorial is to present an elementary analysis of cavity back-action using the classical theories of optics and electrodynamics.

With reference to Fig.1, we analyze the light reflected from a Fabry-Perot cavity consisting of a front (dielectric) mirror having Fresnel reflection and transmission coefficients $\rho$ and $\tau$, and a rear mirror having reflection coefficient $\rho_m$. For the sake of simplicity, we assume that the front mirror is symmetric, meaning that $\rho$ and $\tau$ are the same for incidence from either side [18,19]. We also assume that $\rho$ is real and positive, which implies that $\tau$ is purely imaginary. (There always exists a 90° phase difference between $\rho$ and $\tau$ [18].) Moreover, since the front mirror is lossless, its reflectivity and transmissivity add up to unity, that is, $|\rho|^2 + |\tau|^2 = 1$. Under these circumstances, we will have $\rho^2 - \tau^2 = 1$.

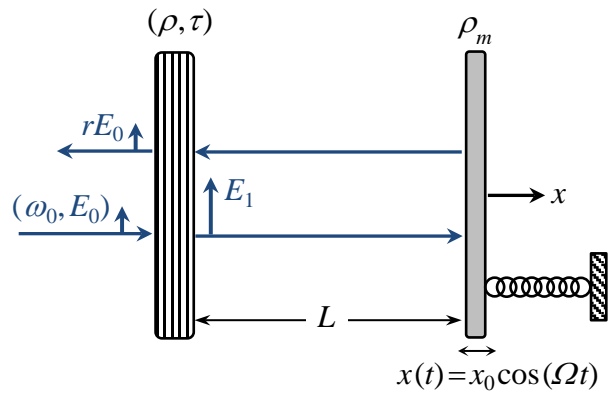

**Fig. 1**. Optical cavity consisting of a dielectric mirror having reflection and transmission coefficients $\rho$ and $\tau$, and a partially reflecting mirror having reflection coefficient $\rho_m$. The two mirrors are separated by an air-gap of width $L$. The rear mirror oscillates along the $x$-axis with frequency $\Omega$ and amplitude $x_0$. The incident beam has amplitude $E_0$ and frequency $\omega_0$. The amplitude of the reflected $E$-field is $rE_0$, while that of the circulating field inside the cavity immediately after the entrance mirror is $E_1$.

The rear mirror's reflection coefficient $\rho_m$ is also real and positive; its average distance from the front mirror is $L$; its location oscillates around an average position by $x(t) = x_0 \cos(\Omega t)$. The normally incident, monochromatic plane-wave has frequency $\omega_0$ and amplitude $E_0$, the reflection coefficient of the Fabry-Perot resonator at its front facet is $r$, and the $E$-field amplitude inside the cavity, immediately after the front mirror and propagating to the right, is $E_1$.

Consider the $E$-field amplitude inside the cavity and immediately after the front mirror. In the absence of mirror vibrations, the self-consistency equation for $E_1$ may be written as follows:

$$E_1 = \tau E_0 + \rho \rho_m \exp(\mathrm{i}2L\omega_0/c)E_1 \quad \rightarrow \quad \frac{E_1}{E_0} = \frac{\tau}{1 - \rho \rho_m \exp(\mathrm{i}2L\omega_0/c)}. \tag{1}$$

The cavity would be on resonance at the incidence frequency $\omega_0$ if the phase of the term in the denominator is an integer-multiple of $2\pi$, that is, $L = m\lambda_0/2$, where $m$ is an integer and $\lambda_0 = 2\pi c/\omega_0$ is the vacuum wavelength [19]. (We are assuming here that $\rho$ and $\rho_m$ are real and positive, otherwise their phase angles will have to be taken into account.) The free spectral range, i.e., the distance between adjacent resonances of the cavity, is $\Delta\omega = \pi c/L$. Supposing that $\rho \rho_m$ is close to unity, the resonance line-width is seen from Eq.(1) to be

$$\delta\omega \approx (c/L)(1 - \rho \rho_m)/\sqrt{\rho \rho_m}. \tag{2}$$

Since $L/c = m\pi/\omega_0$, the cavity quality-factor will be

$$Q_0 = \omega_0/\delta\omega = m\pi\sqrt{\rho \rho_m}/(1 - \rho \rho_m). \tag{3}$$

The finesse $\mathcal{F}$ of the Fabry-Perot resonator, defined as the ratio of the free spectral range $\Delta\omega$ to the line width $\delta\omega$, is similar to the quality-factor given by Eq.(3) except for the integer $m$ being removed. Consequently,

$$\mathcal{F} = \Delta\omega/\delta\omega = \pi\sqrt{\rho \rho_m}/(1 - \rho \rho_m). \tag{4}$$

In the absence of mirror vibrations, the reflection coefficient $r$ at the front facet of the resonator satisfies the following identity:

$$rE_0 = \rho E_0 + \tau \rho_m \exp(\mathrm{i}\,2L\omega_0/c)E_1. \tag{5}$$

Therefore,

$$r = \rho + \frac{\tau^2 \rho_m \exp(\mathrm{i}2L\omega_0/c)}{1 - \rho \rho_m \exp(\mathrm{i}2L\omega_0/c)} = \frac{\rho - (\rho^2 - \tau^2)\rho_m \exp(\mathrm{i}2L\omega_0/c)}{1 - \rho \rho_m \exp(\mathrm{i}2L\omega_0/c)}$$

$$= \frac{\rho - \rho_m \exp(\mathrm{i}2L\omega_0/c)}{1 - \rho \rho_m \exp(\mathrm{i}2L\omega_0/c)}. \tag{6}$$

Having derived from first principles the basic characteristics of the Fabry-Perot resonator of Fig.1 (when the reflector at the rear of the cavity is stationary), we are now in a position to analyze the behavior of the resonator when its rear mirror oscillates with amplitude $x_0$ at frequency $\Omega$.

**2. Adiabatic regime**. In the adiabatic approximation, the cavity length changes slowly enough that $L$ in the preceding equations may be replaced by $L + x(t)$. (In the adiabatic regime the cavity photon lifetime is short compared to the oscillation period $2\pi/\Omega$ of the mirror.) We may then write Eq.(1) as follows:

$$\frac{E_1}{E_0} = \frac{\tau}{1 - \rho \rho_m \exp\{i2(\omega_0/c)[L + x_0\cos(\Omega t)]\}}$$

Geometric series:
$$\sum_{\ell=0}^{\infty} \eta^\ell = \frac{1}{1-\eta} \longrightarrow = \tau\left\{1 + \sum_{\ell=1}^{\infty}\left[\rho\rho_m\exp(i2L\omega_0/c)\right]^\ell\exp[i2\ell(\omega_0/c)x_0\cos(\Omega t)]\right\}. \qquad (7)$$

The function $\exp[i2\ell(\omega_0/c)x_0\cos(\Omega t)]$ appearing in Eq.(7) is a periodic function of time, which may be expanded in a Fourier series, namely,

$$\exp[i2\ell(\omega_0/c)x_0\cos(\Omega t)] = \sum_{n=-\infty}^{\infty}a_n\exp(in\Omega t), \qquad (8a)$$

where

$$a_n = \frac{\Omega}{2\pi}\int_0^{2\pi/\Omega}\exp[i2\ell(\omega_0/c)x_0\cos(\Omega t)]\exp(-in\Omega t)\,dt$$

$$= \frac{1}{2\pi}\int_0^{2\pi}\exp\{i[2\ell(\omega_0/c)x_0\cos x - nx]\}\,dx = i^n J_n(2\ell\omega_0 x_0/c). \quad \longleftarrow \boxed{\text{See [20], 8.411-1}} \quad (8b)$$

In the above equation, $J_n(\cdot)$ is a Bessel function of the first kind, $n^{\text{th}}$ order. Equation (7) may now be written

$$\frac{E_1}{E_0} = \tau\left\{1 + \sum_{n=-\infty}^{\infty}i^n\left[\sum_{\ell=1}^{\infty}\rho^\ell\rho_m^\ell\exp(i2\ell\omega_0 L/c)\,J_n(2\ell\omega_0 x_0/c)\right]\exp(in\Omega t)\right\}. \qquad (9)$$

The cavity field has thus acquired sidebands at frequencies $\omega_0 \pm n\Omega$. Note, however, that the field amplitudes for $+n$ and $-n$ sidebands are identical, because $J_{-n}(\cdot) = (-1)^n J_n(\cdot)$ and $i^{-n} = (-i)^n$. This sideband symmetry is a direct consequence of the adiabatic approximation; it will go away when the photon lifetime becomes comparable to or longer than the oscillation period of the mirror, as will be seen in the next section.

The reflection coefficient $r$ at the cavity's front facet may likewise be expressed in the adiabatic approximation as a superposition of terms associated with the sidebands. We will have

$$r = \frac{\rho - \rho_m\exp\{i2(\omega_0/c)[L + x_0\cos(\Omega t)]\}}{1 - \rho\rho_m\exp\{i2(\omega_0/c)[L + x_0\cos(\Omega t)]\}}$$

$$= \{\rho - \rho_m\exp\{i2(\omega_0/c)[L + x_0\cos(\Omega t)]\}\}\sum_{\ell=0}^{\infty}\left[\rho\rho_m\exp(i2\omega_0 L/c)\right]^\ell\exp[i2\ell(\omega_0/c)x_0\cos(\Omega t)]$$

$$= \rho + \sum_{\ell=1}^{\infty}(\rho^{\ell+1} - \rho^{\ell-1})\rho_m^\ell\exp(i2\ell\omega_0 L/c)\exp[i2\ell(\omega_0/c)x_0\cos(\Omega t)]$$

$$= \rho + (\rho - \rho^{-1})\sum_{n=-\infty}^{\infty}i^n\left[\sum_{\ell=1}^{\infty}\rho^\ell\rho_m^\ell\exp(i2\ell\omega_0 L/c)\,J_n(2\ell\omega_0 x_0/c)\right]\exp(in\Omega t). \qquad (10)$$

Again, the reflected light is seen to have sidebands at frequencies $\omega_0 \pm n\Omega$ and, in this adiabatic regime, the $+n$ and $-n$ sideband amplitudes are identical.

**3. Beyond adiabatic approximation**. If the oscillation period $2\pi/\Omega$ of the rear mirror happens to be short compared to the cavity photon lifetime, we need to approach the problem differently. Let the light amplitude incident on the rear mirror be $E_1 \exp(\mathrm{i}L\omega_0/c)\exp(-\mathrm{i}\omega_0 t)$. The vibrations of the mirror modulate the phase of the incident beam so that the reflected $E$-field at the equilibrium location of the mirror may be written as follows:

$$E_{\mathrm{ref}}(x=L,t) \approx \rho_m E_1 \exp\{\mathrm{i}(\omega_0/c)[L+2x_0\cos(\Omega t)]\}\exp(-\mathrm{i}\omega_0 t). \tag{11}$$

Several approximations are involved in writing the above equation, although, for small values of $x_0$, the resulting errors are relatively minor and may be readily ignored. Using the Fourier series expansion

$$\exp[\mathrm{i}2(\omega_0/c)x_0\cos(\Omega t)] \;=\; \sum_{n=-\infty}^{\infty}\mathrm{i}^n J_n(2\omega_0 x_0/c)\exp(\mathrm{i}n\Omega t), \tag{12}$$

we rewrite Eq.(11) as follows:

$$E_{\mathrm{ref}}(x=L,t) \approx \rho_m E_1 \exp(\mathrm{i}L\omega_0/c)\sum_{n=-\infty}^{\infty}\mathrm{i}^n J_n(2\omega_0 x_0/c)\exp[-\mathrm{i}(\omega_0-n\Omega)t]. \tag{13}$$

It should now be obvious that the vibrating mirror acts as a scatterer, throwing a fraction of the incident light into the sideband frequencies $\omega_0 \pm n\Omega$. The effective reflection coefficient of the mirror for the incident frequency $\omega_0$ now becomes $\rho_m J_0(2\omega_0 x_0/c)$, requiring this change to be incorporated into Eqs.(1-6) to yield the correct values of $E_1$, $\delta\omega$, $Q_0$, $\mathcal{F}$, and $r$ for the incidence frequency $\omega_0$.

For $n \geq 2$, the light scattered into $\omega_0 \pm n\Omega$ is too weak to concern us here. We thus focus our attention on the first-order terms with frequencies $\omega_0 \pm \Omega$ (often referred to as Stokes and anti-Stokes terms), which enter the cavity with equal amplitudes $\mathrm{i}\rho_m J_1(2\omega_0 x_0/c)\exp(\mathrm{i}L\omega_0/c)E_1$. The cavity then builds these sideband frequencies into cavity modes having the following amplitudes at the equilibrium position of the mirror:

$$\begin{aligned}
E^{(\mathrm{cavity})}_{\omega_0\pm\Omega} &= \frac{\mathrm{i}\rho_m J_1(2\omega_0 x_0/c)\exp(\mathrm{i}L\omega_0/c)E_1}{1-\rho\rho_m\exp[\mathrm{i}2L(\omega_0\pm\Omega)/c]}\\[2mm]
&= \frac{\mathrm{i}\tau\rho_m J_1(2\omega_0 x_0/c)\exp(\mathrm{i}L\omega_0/c)E_0}{[1-\rho\rho_m J_0(2\omega_0 x_0/c)\exp(\mathrm{i}2L\omega_0/c)]\{1-\rho\rho_m\exp[\mathrm{i}2L(\omega_0\pm\Omega)/c]\}}.
\end{aligned} \tag{14}$$

Subsequently, the cavity sidebands emerge from the front mirror with the following amplitudes:

$$\frac{E^{(\mathrm{emergent})}_{\omega_0\pm\Omega}}{E_0} = \frac{\mathrm{i}\tau^2\rho_m J_1(2\omega_0 x_0/c)\exp(\mathrm{i}2L\omega_0/c)\exp(\pm\mathrm{i}L\Omega/c)}{[1-\rho\rho_m J_0(2\omega_0 x_0/c)\exp(\mathrm{i}2L\omega_0/c)]\{1-\rho\rho_m\exp[\mathrm{i}2L(\omega_0\pm\Omega)/c]\}}. \tag{15}$$

Note that the second bracketed term in the denominator of Eq.(15) can give rise to asymmetry between the Stokes and anti-Stokes sidebands. In particular, if the incident frequency $\omega_0$ is slightly detuned from resonance, one of the sidebands moves closer to resonance while the other one moves further away, thus ensuring that one emergent sideband is stronger than the other. Suppose, for instance, that the incident beam is red-detuned. The blue-shifted anti-Stokes sideband (frequency $= \omega_0 + \Omega$) is then stronger than the red-shifted Stokes sideband (frequency $= \omega_0 - \Omega$). Since blue-shifted photons are associated with reflection from a mirror moving toward the source — whereas red-shifted light represents reflection from a receding mirror — it is clear that the back-action of cavity photons under the above circumstances must slow down the oscillating mirror. This is the principle of cooling a vibrating mirror by the back-action of cavity photons, which will be further elaborated below.

The following analysis will be substantially simplified if we introduce some new parameters and make a few approximations at this point. For small values of $x_0$ we may write $J_0(2\omega_0 x_0/c) \approx 1$ and $J_1(2\omega_0 x_0/c) \approx \omega_0 x_0/c$ [20]. Let $\omega_r$ be the closest resonance frequency to the incidence frequency, $\omega_0$, and assume that $\omega_0$ as well as $\omega_0 \pm \Omega$ are sufficiently close to resonance that we may write

$$\exp(\mathrm{i}2L\omega_0/c) = \exp[\mathrm{i}2L(\omega_0 - \omega_r)/c] \approx 1 - \mathrm{i}2L(\omega_r - \omega_0)/c, \quad \boxed{\exp(\mathrm{i}2L\omega_r/c) = 1} \quad (16a)$$

$$\exp[\mathrm{i}2L(\omega_0 \pm \Omega)/c] = \exp[\mathrm{i}2L(\omega_0 \pm \Omega - \omega_r)/c] \approx 1 - \mathrm{i}2L(\omega_r - \omega_0 \mp \Omega)/c. \quad (16b)$$

For the terms in the denominator of Eq.(1) and Eq.(14), we assume that $\rho \rho_m$ is sufficiently close to unity that we may replace $1 - \rho\rho_m$ with $(L/c)\delta\omega$ [see Eq.(2)], and set $\rho\rho_m = 1$ for the remaining coefficients. At the equilibrium position of the rear mirror, we will then have

$$E_{\omega_0}^{(\text{cavity})} = \frac{\tau \exp(\mathrm{i}L\omega_0/c)E_0}{1 - \rho\rho_m J_0(2\omega_0 x_0/c)\exp(\mathrm{i}2L\omega_0/c)} \approx \frac{\tau \exp(\mathrm{i}L\omega_0/c)E_0}{(L/c)[\delta\omega + \mathrm{i}2(\omega_r - \omega_0)]}. \quad (17)$$

$$E_{\omega_0 \pm \Omega}^{(\text{cavity})} \approx \frac{\mathrm{i}\tau(\omega_0 x_0/c)\exp(\mathrm{i}L\omega_0/c)E_0}{(L/c)^2[\delta\omega + \mathrm{i}2(\omega_r - \omega_0)][\delta\omega + \mathrm{i}2(\omega_r - \omega_0 \mp \Omega)]}$$

$$= \frac{\mathrm{i}\tau(\mathcal{F}/\pi)^2(\omega_0 x_0/c)\exp(\mathrm{i}L\omega_0/c)E_0}{[1 + \mathrm{i}2(\omega_r - \omega_0)/\delta\omega][1 + \mathrm{i}2(\omega_r - \omega_0 \mp \Omega)/\delta\omega]}. \quad \boxed{\text{See Eqs.(2) and (4)}} \quad (18)$$

The total $E$-field incident on the rear mirror may thus be written as follows:

$$E_{\text{total}}^{(\text{cavity})} \approx \frac{\tau \exp(\mathrm{i}L\omega_0/c)E_0\exp(-\mathrm{i}\omega_0 t)}{(L/c)[\delta\omega + \mathrm{i}2(\omega_r - \omega_0)]} \left\{ 1 + \frac{\mathrm{i}(L/c)\delta\omega(\mathcal{F}/\pi)^2(\omega_0 x_0/c)}{1 + \mathrm{i}2(\omega_r - \omega_0 - \Omega)/\delta\omega}\exp(-\mathrm{i}\Omega t) \right.$$

$$\left. + \frac{\mathrm{i}(L/c)\delta\omega(\mathcal{F}/\pi)^2(\omega_0 x_0/c)}{1 + \mathrm{i}2(\omega_r - \omega_0 + \Omega)/\delta\omega}\exp(\mathrm{i}\Omega t) \right\}. \quad (19)$$

To find the radiation pressure on the vibrating mirror, we need to calculate the time-averaged Poynting vector $\langle \boldsymbol{S} \rangle$ of the incident field [21-24]. Considering that a plane-wave's magnetic field amplitude $H_0$ is related to its electric field amplitude $E_0$ via $H_0 = E_0/Z_0$, where $Z_0 = \sqrt{\mu_0/\varepsilon_0} \approx 377$ ohm is the impedance of free space ($\mu_0$ and $\varepsilon_0$ being the permeability and permittivity of free space) [21], we will have

$$<S_x> = \tfrac{1}{2}\mathrm{Re}(\boldsymbol{E} \times \boldsymbol{H}^*)_x = \frac{|\tau|^2|E_0|^2}{2Z_0(L/c)^2\delta^2\omega[1 + 4(\omega_r - \omega_0)^2/\delta^2\omega]}$$

$$\times \left\{ 1 + \frac{(L/c)^2\delta^2\omega(\mathcal{F}/\pi)^4(\omega_0 x_0/c)^2}{1 + 4(\omega_r - \omega_0 - \Omega)^2/\delta^2\omega} + \frac{(L/c)^2\delta^2\omega(\mathcal{F}/\pi)^4(\omega_0 x_0/c)^2}{1 + 4(\omega_r - \omega_0 + \Omega)^2/\delta^2\omega} \right. \quad \boxed{\begin{array}{c}\text{Continued on}\\\text{the next page}\end{array}}$$

$$+ \frac{2(L/c)\delta\omega(\mathcal{F}/\pi)^2(\omega_0 x_0/c)\left\{2[(\omega_r - \omega_0 - \Omega)/\delta\omega]\cos(\Omega t) + \sin(\Omega t)\right\}}{1 + 4(\omega_r - \omega_0 - \Omega)^2/\delta^2\omega}$$

$$+ \frac{2(L/c)\delta\omega(\mathcal{F}/\pi)^2(\omega_0 x_0/c)\left\{2[(\omega_r - \omega_0 + \Omega)/\delta\omega]\cos(\Omega t) - \sin(\Omega t)\right\}}{1 + 4(\omega_r - \omega_0 + \Omega)^2/\delta^2\omega}$$

$$+ \frac{2(L/c)^2\delta^2\omega(\mathcal{F}/\pi)^4(\omega_0 x_0/c)^2\left\{\{1 + 4[(\omega_r - \omega_0)^2 - \Omega^2]/\delta^2\omega\}\cos(2\Omega t) + 4(\Omega/\delta\omega)\sin(2\Omega t)\right\}}{[1 + 4(\omega_r - \omega_0 - \Omega)^2/\delta^2\omega][1 + 4(\omega_r - \omega_0 + \Omega)^2/\delta^2\omega]}\Bigg\}.$$

$$(20)$$

To find the radiation pressure on the vibrating mirror, we divide the above expression for $\langle S_x \rangle$ (i.e., the time-averaged component of the Poynting vector along $x$) by $c$, the speed of light in vacuum, then multiply by 2 to account for the reversal of momentum upon reflection [21-24]. (Note that the power of the reflected beam from the vibrating mirror is essentially the same as that of the incident beam. One way to see this is to look at the opposite end of the cavity and to observe that $\rho \approx 1$; the round-trip phase, of course, is irrelevant here.) Let us denote by $P_{\text{in}} = \frac{1}{2}|E_0|^2 A/Z_0$ the incident optical power on the area $A$ of the mirror at the front facet of the cavity. The first three terms in Eq.(20) are time-independent; they compress the spring attached to the mirror by a fixed amount, which affects neither the spring constant nor the damping coefficient. As for the last term in Eq.(20), which has twice the vibration frequency $\Omega$, its magnitude is probably too small to be consequential. This leaves the 4th and 5th terms in Eq.(20), which yield the effective force $F_{\text{RP}}$ of radiation pressure on the vibrating mirror as follows:

$$F_{\text{RP}} = \frac{4|\tau|^2(\mathcal{F}/\pi)^2\omega_0 P_{\text{in}}}{c^2(L/c)\delta\omega[1 + 4(\omega_r - \omega_0)^2/\delta^2\omega]}$$

$$\times \left\{\frac{2[(\omega_r - \omega_0 - \Omega)/\delta\omega]x_0\cos(\Omega t) + x_0\sin(\Omega t)}{1 + 4(\omega_r - \omega_0 - \Omega)^2/\delta^2\omega} + \frac{2[(\omega_r - \omega_0 + \Omega)/\delta\omega]x_0\cos(\Omega t) - x_0\sin(\Omega t)}{1 + 4(\omega_r - \omega_0 + \Omega)^2/\delta^2\omega}\right\}$$

$$(21)$$

In the above equation, the terms that are proportional to the mirror's position, $x_0\cos(\Omega t)$, modify the spring constant $\alpha$ and, therefore, the resonance frequency of the mirror, $\Omega = \sqrt{\alpha/m_{\text{eff}}}$; here $m_{\text{eff}}$ is the mirror's effective mass [23,24]. For small changes of $\alpha$, the change in the resonance frequency is thus given by $\Delta\Omega = \Delta\alpha/(2m_{\text{eff}}\,\Omega)$. We have

$$\Delta\Omega = -\frac{4|\tau|^2(\mathcal{F}/\pi)^2\omega_0 P_{\text{in}}}{m_{\text{eff}}\,c^2(L/c)\delta^2\omega[1 + 4(\omega_r - \omega_0)^2/\delta^2\omega]\,\Omega}$$

$$\times \left[\frac{\omega_r - \omega_0 - \Omega}{1 + 4(\omega_r - \omega_0 - \Omega)^2/\delta^2\omega} + \frac{\omega_r - \omega_0 + \Omega}{1 + 4(\omega_r - \omega_0 + \Omega)^2/\delta^2\omega}\right].$$

$$(22)$$

As for the terms appearing in Eq.(21) which are proportional to the mirror velocity, $dx/dt = -x_0\Omega\sin(\Omega t)$, they represent a contribution to the friction coefficient $\beta$ of the spring in the form of $F_{RP} = -\beta\,dx/dt$ [23,24]. The corresponding contribution to the damping coefficient $\gamma$ of the mirror is thus given by

$$\gamma_{\text{RP}} = \beta/m_{\text{eff}} = \frac{4|\tau|^2 (\mathscr{F}/\pi)^2 \omega_0 P_{\text{in}}}{m_{\text{eff}} c^2 (L/c) \, \delta\omega \left[1 + 4(\omega_r - \omega_0)^2/\delta^2\omega\right] \Omega}$$

$$\times \left[ \frac{1}{1 + 4(\omega_r - \omega_0 - \Omega)^2/\delta^2\omega} - \frac{1}{1 + 4(\omega_r - \omega_0 + \Omega)^2/\delta^2\omega} \right]. \tag{23}$$

In this way, the damping effect of the field trapped in the cavity on the vibrating mirror is quantified. Note, in particular, that in the absence of detuning, i.e., when $\omega_0 = \omega_r$, radiation-pressure-induced damping ceases to exist.